\documentclass[10pt,nofootinbib,amsmath,amssymb,showpacs,twocolumn]{revtex4-1}
\usepackage[T1]{fontenc} \usepackage[latin1]{inputenc}
\usepackage{amsmath,color} 
\usepackage{graphicx}
\usepackage{epsfig} \usepackage{amssymb} 
\usepackage[capitalize]{cleveref}

\newcommand{\vdwfm}{Van der Waals ferromagnets}

\begin{document}

\title{Layer effects on the magnetic textures in magnets with local inversion asymmetry}
\author{E. van Walsem$^1$}
\author{R. A. Duine$^{1,2}$}
\author{M. H. D. Guimar\~aes$^{2,3}$}
\affiliation{$^1$Institute for Theoretical Physics, Universiteit Utrecht, Leuvenlaan 4, 3584 CE Utrecht, The Netherlands}
\affiliation{$^2$ Department of Applied Physics, Eindhoven University of Technology, P.O. Box 513, 5600 MB Eindhoven, The Netherlands}
\affiliation{$^3$ Zernike Institute for Advanced Materials, University of Groningen, Groningen, the Netherlands.  }

\pacs{PACS NRS}

\begin{abstract}
Magnets with broken local inversion symmetries are interesting candidates for chiral magnetic textures such as skyrmions and spin spirals. 
The property of these magnets is that each subsequent layer can possess a different Dzyaloshniskii-Moriya interaction (DMI) originating from the local inversion symmetry breaking.
Given that new candidates of such systems are emerging, with the Van der Waals crystals and magnetic multilayer systems, it is interesting to investigate how the chiral magnetic textures depend on the number of layers and the coupling between them.
In this article, we model the magnetic layers with a classical Heisenberg spin model, where the sign of the DMI alternates for each consecutive layer. We use Monte Carlo simulations to examine chiral magnetic textures and show that the pitch of magnetic spirals is influenced by the interlayer coupling and the number of layers. We observe even-odd effects in the number of layers, where we observe a suppression of the spin spirals for even layer numbers. We give an explanation for our findings by proposing a net DMI in systems with strongly coupled layers. Our results can be used to determine the DMI in systems with a known number of layers, and for new technologies based on the tunability of the spiral wave length.
\end{abstract}

\maketitle 

\section{introduction}
The role of electronic devices in society is ever increasing, and there is a need to make them smaller and faster, while keeping a low power consumption. However, the current technologies are reaching their limits since the information density cannot be increased much further. 
Therefore, new technologies need to be developed. One of the most promising new technologies for data processing and storage are magnetic systems with chiral textures such as chiral magnetic domain walls and skyrmions \cite{Bode2007,Heide2008,Thiaville2012,Skyrme1962,Bogdanov1994}. Their chiral and topological properties make for sturdy textures which can become extremely small, this makes them suitable for applications.  
An example of such a design is the skyrmion-racetrack memory, which is a promising route for fast and energy efficient memory and processing devices \cite{Parkin2008,Kiselev2011,Fert2013}. 

\begin{figure}[h!]
\includegraphics[width=0.65\linewidth]{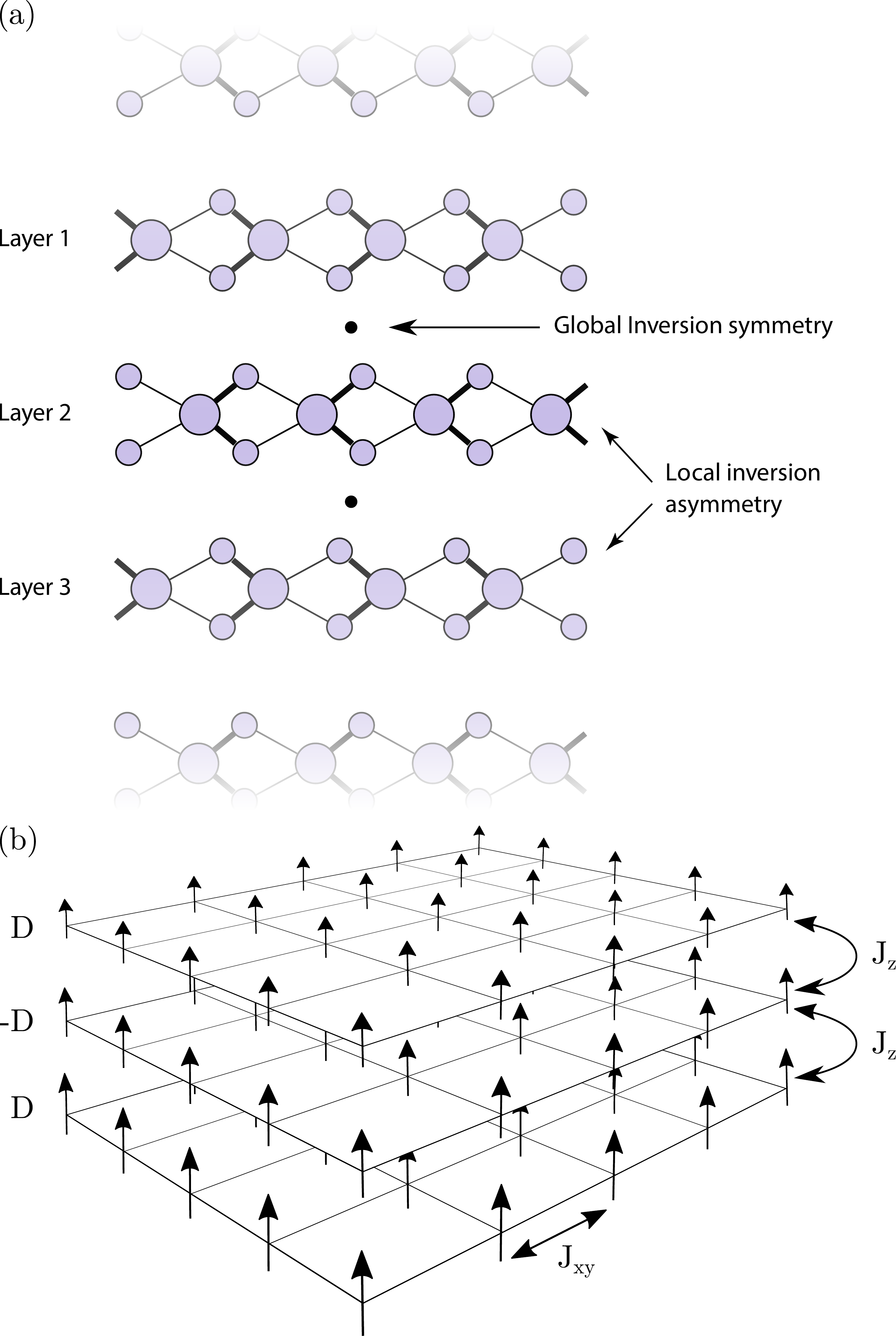}
\caption{Schematics of (a) a bulk crystal with a global inversion symmetry while showing local inversion asymmetry, and (b) the model used in our simulations. Here, the arrows are Heisenberg spins and can rotate freely in three dimensions, $J_{\rm xy}$ is the intralayer coupling, $J_{\rm z}$ the interlayer coupling and D the DMI.}
\label{fig:model}
\end{figure}

At the moment, a plethora of systems is known to host skyrmionic textures such as the chiral magnet MnSi which hosts so called Bloch skyrmions and ultra-thin ferromagnetic films which typically host N\'eel skyrmions \cite{Muhlbauer2009,Yu2010,Heinze2011,Dupe2014,Yu2016,Woo2016,Soumyanarayanan2017,Moreau-Luchaire2016,Chen2015}. 
These thin magnetic films are formed by stacking multiple layers of different metals and the order of these layers determine their magnetic properties.
Furthermore, a new class of suitable materials is emerging: the Van der Waals crystals. These crystals consists of two dimensional layers of one atom thick, stacked on top of each other via Van der Waals bonds \cite{Geim2013,Gibertini2019}. Two-dimensional layers can be exfoliated from bulk materials such as graphite, hexagonal boron nitride (hBN) and $\rm{CrI_3}$ \cite{Novoselov2004,Geim2013,Huang2017}. 
Because of the freedom to stack different kind of materials, the end product is tunable and can be formed such that the desired properties are present in the end product with virtually no strain since the weak interlayer bonds make them less sensitive to lattice mismatch problems.

\begin{figure*} 
\includegraphics[width=\linewidth]{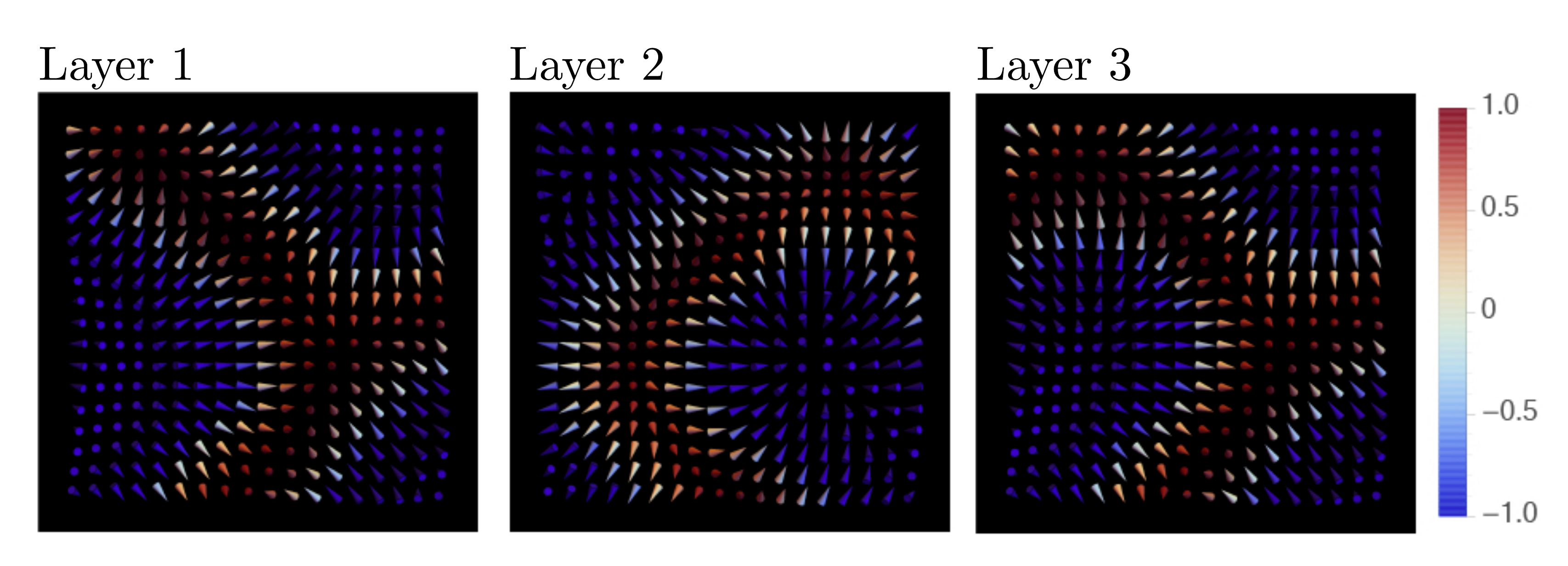}
\caption{Snapshot of a 32x32 cut out of a three layer 128x128 spinsystem with $J_{\rm z}/J_{\rm xy}=0.10$ and $D/J_{\rm xy}=0.50$. The color red indicates that a spin is pointing upwards and blue downwards. Spin spirals are clearly visible, and the sign of the DMI has a clear effect on the turning sense (anti-clockwise for layer 1 and 3, and clockwise for layer 2). The maze-like structure in the spin spirals are formed because of the coupling between the layers. }
\label{fig:snapshot}
\end{figure*}

In this article we demonstrate how a local DMI arising from a local inversion asymmetry can give rise to chiral structures even in materials with global inversion symmetry. DMI is an interaction formed when strong spin orbit coupling and broken inversion symmetry are present. An example of such symmetry breaking is the interface between two different materials such as Co and Pt \cite{Dzyaloshinsky1958,moriya1960}. The DMI is also referred to as antisymmetric exchange since the interaction picks up a minus sign when exchanging two spins. Because of this property, the interaction leads to chiral magnetic textures. One example of chiral magnetic textures are spirals with a preferred turning sense (either clockwise or counterclockwise) which are formed by DMI in the absence of fields and anisotropies. 
Here we note that a local DMI can arise in crystal possessing global inversion symmetry, but which show a local inversion asymmetry (as drawn in \cref{fig:model}). This allows for a DMI term to be non-zero locally while averaging out when the complete infinite crystal is taken into account
This is analogous to the ``hidden spin polarization'' effect that occurs because of local inversion symmetry breaking and was elucidated in Ref. \cite{Zhang2014} and experimentally verified in Van der Waals crystals \cite{Riley2014,Bertoni2016,Razzoli2017,Guimaraes2018}. We are interested in magnets where this local DMI has an alternating nature of its sign in subsequent layers. 
Especially, we are interested in materials with (anti)ferromagnetic coupling between the layers. 
To comply with this condition Van der Waals crystals need to have bulk inversion symmetry, layer inversion asymmetry, magnetism and high spin orbit coupling. 
An example of a Van der Waals material meeting these criteria is FGT \cite{Gibertini2019}, which belongs to the space group $P6_{3}/mmc$ \cite{Kim2018} in its bulk form and point group $D_{3h}$ in its monolayer form \cite{Johansen2019}. 
The condition of local inversion asymmetry in a globally inversion symmetric system can also be obtained in sputtered metallic thin films, such as Ta/Co/Pt/Co/Ta systems making it even easier to perform such DMI engineering \cite{Lucassen2020}.

In this article we discuss how a locally nonzero DMI influences magnetic textures. We show what different textures form and find that these chiral textures occurring in the system are influenced by the stacking of the layers. The resulting spin spiral wavelength and their turning sense are affected by the interlayer coupling relative to the DMI. Furthermore, we find that the number of layers influences the wavelength of the spin spirals in the system, and an even-odd effect is found for the number of layers in the system. 
The tunability of the spin spirals suggests that skyrmions will also be tunable in their size. 
The tunability of the spin spiral wavelength and turning sense demonstrate the potential of DMI engineering for new magnetic devices. Moreover, since spin spiral-systems can develop skyrmions upon applied magnetic fields, our results also serve as a basis for the design of skyrmionic devices.

The remainder of this paper is organized as follows. In \cref{secMethod} we discuss our model and how the simulations are performed. After this we show results of our simulations in  \cref{secResults} where we focus on the phase diagram, wavelength and turning sense found in the system with an odd number of layers. This is compared in \cref{secEven} to systems with an even number of layers. Finally, we conclude with an outlook in \cref{secConcl}.

\section{Model and Method} \label{secMethod}
To model a stack of coupled ferromagnetic layers, we describe them with a classical Heisenberg spin model. Each layer is modeled by equally spaced spins on a square lattice. This is done for simplicity and is a decent approximation since we are modeling temperatures far below the Curie temperature and are interested in smooth textures such as spin spirals. The layers are placed right on top of each other, as shown in \cref{fig:model} (b), and each layer has an alternating sign for the DMI strength	.
We assume that the leading interactions within the layers are ferromagnetic nearest neighbour exchange, and DMI. 
The leading interaction between the layers is assumed to be nearest neighbour exchange varying from the ferromagnetic to the antiferromagnetic regime. We note that these conditions are met for various Van der Waals crystals as well as for metallic thin film heterostructures with RKKY coupled layers.

We describe the hamiltonian, $H$, with separate terms for the intra- and inter-layer terms:
\begin{equation}
H = H_{\rm intra} + H_{\rm inter}.
\end{equation}
The interactions within the layers are expressed as follows: 
\begin{align}\label{eq:hamiltonian}
H_{\rm intra} =
	-& J_{\rm xy} \sum_{\alpha=1}^{N} \sum_{\bf r}  {\bf S}^{\alpha}_{\bf r}  \cdot \left(  {\bf S}^{\alpha}_{{\bf r}+{\bf \hat{x}}} +  {\bf S}^{\alpha}_{{\bf r}+{\bf \hat{y}}} \right) + \nonumber \\ 
	 \sum_{\alpha=1}^{N}  (-1)^{\alpha-1}& \cdot D\sum_{\bf r} \left(  {\bf S}^{\alpha}_{\bf r} \times {\bf S}^{\alpha}_{{\bf r}+{\bf \hat{x}}} \cdot {\bf \hat{y}} - {\bf S}^{\alpha}_{\bf r} \times {\bf S}^{\alpha}_{{\bf r}+{\bf \hat{y}}} \cdot {\bf \hat{x}} \right),  
\end{align}
where $J_{\rm xy}$  is the intralayer coupling, $D$ is the Dzyaloshinskii-Moriya interaction, $ {\bf S}^{\alpha}_{\bf r} $ is the spin at position ${\bf r}$ in layer $\alpha = 1,2,...$ and $\bf{\hat{x}}$ and $\bf{\hat{y}}$ are the unit vectors in the x and y direction respectively. The number of layers is denoted with $N$ and the $(-1)^\alpha$ term regulates the alternating DMI sign in the system. The interactions between the layers are described by:
\begin{align}
H_{\rm inter} = 	-& J_{\rm z} \sum_{\alpha=1}^{N} \sum_{\bf r}  {\bf S}^{\alpha}_{\bf r}  \cdot  {\bf S}^{\alpha+1}_{\bf r} ,\nonumber \\ 
\end{align}
where $J_{\rm z}$ is the intralayer coupling.

To investigate the ground state of this model we use Monte Carlo simulations. We begin the simulation by taking a random spin configuration at a high temperature. Then we use the Metropolis algorithm to thermalize the system \cite{Metropolis1953,Hastings1970}. This algorithm picks a random spin and proposes a new semi-random direction. This new direction is such that the average acceptance ratio is 50\%, which is determined from the energy difference between the new and the old spin configuration: $\Delta E$. The acceptance probability $\mathcal{P}$ is then $\mathcal{P}=\exp(\Delta E/k_{\rm B}T) $ if $\Delta E < 0$ or $\mathcal{P} = 1$ in all other cases. 
Finally, we accept or reject this new direction with a probability of $0.5$.
When the system is fully thermalized we decrease the temperature and repeat this thermalization and lowering of the temperature until the temperature gets close to zero and approaches the ground state.
We start our simulations at $k_{\rm B} T / J_{\rm xy} = 10$, where $k_{\rm B}$ is the Boltzman constant and $T$ is the temperature. To this end, the temperature is lowered by a factor of $0.95$ until  $k_{\rm B} T / J_{\rm xy} = 0.01$.

\section{Results for odd number of layers} \label{secResults}
For low temperatures, we expect that magnetic textures will form in the ferromagnet. The DMI leads to the formation of a chiral spiral inside the layer and, depending on the sign of the DMI, the turning sense of the spiral should be different in each subsequent layer. The size of the spiral should be influenced by the relative strength between the DMI and intralayer coupling terms ($D/J_{\rm xy}$).
We performed simulations for varying parameters of DMI ($D/J_{\rm xy}$) and interlayer coupling ($J_{\rm z}/J_{\rm xy}$) and systems with sizes varying between 32x32 spins to 256x256 spins. We find that the systems thermalize and form a stable state where a clear chiral spiral pattern is visible. In \cref{fig:snapshot} we show snapshots of such a system with chiral spirals. We did simulations for a varying number of layers, and in this part of the article we will discuss systems with an odd number of layers. The even number of layer systems are discussed later in \cref{secEven}. 

\subsection{Phases and Phase Diagram}\label{sec:phasediagram}
The magnetic structures in our systems will be determined by the competition between the three terms in the model: the intralayer coupling, which favors the alignment of the spins inside the layer, the DMI, which leads to a chiral spiral inside of the layer, and the interlayer coupling, which favors the (anti-)alignment of the spins between the layers. 

\begin{figure}
\includegraphics[width=0.9\linewidth]{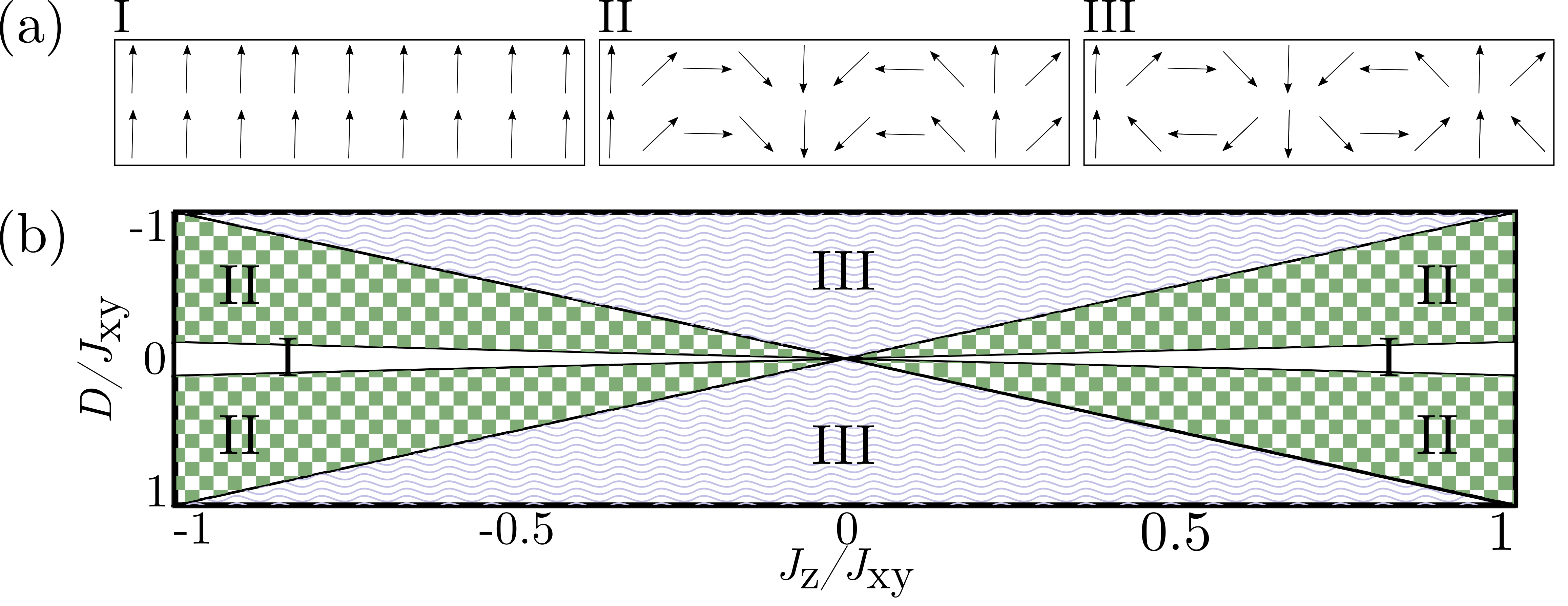}
\caption{(a) Cartoon of two coupled spin layers corresponding to the phases described above. 
(b) Phase diagram of an odd multilayer system. Phase I corresponds to the polarized phase, phase II to the interlayer coupling dominant phase and phase III to the DMI dominant phase. The phase diagram is determined with help from the second layer of a 5 layers system with 256x256 spins with a DMI between 1 and -1, and a $J_{\rm z}/J_{\rm xy}$ between 1 and -1. }
\label{fig:phasediagram}
\end{figure}

Our simulations show three prominent magnetic phases. The first phase is the fully polarized phase where the interlayer exchange is dominant, i.e. $|J_{\rm z}|/J_{\rm xy} \gg D/J_{\rm xy}$. Here no magnetic structure appears, all spins are always aligned. A cartoon of this is shown in \cref{fig:phasediagram} (a) I, and from here on we will refer to this phase as phase I. 
The next phase, phase II, is obtained by increasing the DMI and magnetic spirals start to form. Here the DMI term is non-negligible when compared to $J_{\rm z}$. We have observed this for all finite non zero values of $D/J_{\rm xy}$. Since the interlayer coupling is still fairly large compared to the DMI in this phase, the system behaves like the complete system possesses a single 'net' DMI value and all spirals have the same turning sense, as is shown in \cref{fig:phasediagram} (a) II. Important to note here is that the system we consider in this section is still inversion asymmetric due to the odd number of layers which does allow for a net DMI to be present.
The last phase is where the DMI is dominant and since the sign of the DMI is alternating, the turning sense of the spirals is also alternating in subsequent layers. As is drawn in \cref{fig:phasediagram} (a) III. Here, the interlayer exchange is present as a small perturbation in the system.
The snapshot in \cref{fig:snapshot} is taken in the DMI dominant phase (III), an anti-clockwise turning sense is visible in layers 1 and 3 and a clockwise turning sense in layer 2.
In \cref{fig:phasediagram} (b) we plot the phase diagram where these three phases occur for different values of DMI and interlayer coupling. 

\begin{figure}
\includegraphics[width=0.9\linewidth]{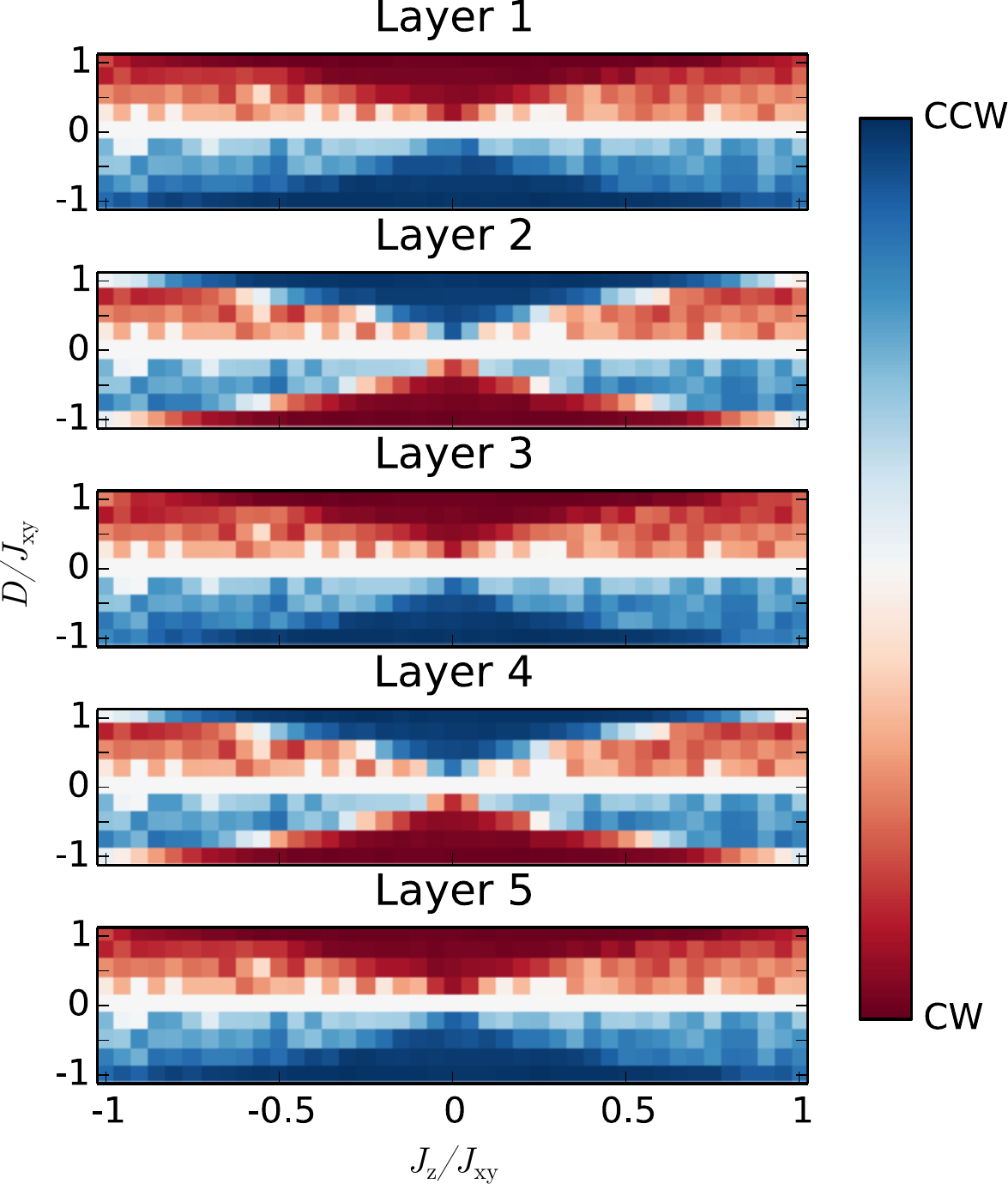}
\caption{Average turning sense in each layer. To calculate this average turning sense each spin is given a value of $\pm1$ corresponding to the orientation with respect to their neighboring spin, finally an average of all these values is calculated. A value of +1 corresponds to a clockwise (CW) turning sense and -1 to an counter clockwise (CCW) turning sense. This is plotted for 5 layers of 256x256 spins with a DMI between 1 and -1, and a $J_{\rm z}/J_{\rm xy}$ between 1 and -1. }
\label{fig:turningsense}
\end{figure}

In \cref{fig:turningsense} we show the average turning sense for a system with 5 layers. Here the fully polarized phase I is clearly visible as the white region where $D/J_{\rm xy}=0$ and no turning sense is present. 
For increased DMI a non-zero average turning sense is visible, corresponding to phase II. In each subsequent layer the turning sense has the same direction. Increasing the DMI even further we find phase III, and also this is clearly visible in \cref{fig:turningsense}. There is a stronger turning sense and each subsequent layer switches the turning sense following the sign of the DMI. In layer 2 and 4, a right (left) turning sense is found for a negative (positive) value of $D/J_{\rm xy}$ instead of a left (right) turning sense.

No clear difference between ferromagnetic ($J_{\rm z} > 0$) and antiferromagnetic ($J_{\rm z} < 0$) interlayer coupling is visible in the phase diagram, i.e. all plots in \cref{fig:turningsense} are symmetric around $J_{\rm z}/J_{\rm xy}$. Examining the snapshots we see that the ferromagnetic coupled layers have aligned spirals and antiferromagnetic coupled layers anti-aligned spirals. In the ferromagnetic case, the spins pointing perpendicular to the layer are aligned between the layers, but due to the alternating nature of the DMI, the parallel pointing spins are anti-aligned. In the antiferromagnetic case the opposite is true: the perpendicular spins are anti-aligned, but the parallel spins are not. This contrast makes that there is no noticeable energy difference between the ferromagnetic and antiferromagnetic case.

\subsection{Wave vector}
In this section we will focus on the wave vector of the spirals found in phase II and III. The wave vector is defined by the number of cycles a spiral forms per spin, and is determined by the ratio of DMI and intralayer exchange coupling. A higher ratio leads to a shorter spiral period and thus a larger wave vector. 
We expect that the competition between the DMI and interlayer coupling has significant effects on the spirals, since two coupled spirals with different turning sense cannot be coupled such that all spins are aligned. In \cref{fig:phasediagram} (a) III it is visible that the interlayer coupling between the two spirals gives a different energy contribution per spin: a favorable energy contribution for the vertically aligned spins, and an unfavorable one for the horizontal anit-aligned spins. It is impossible to shift the spirals relative to each other such that the interlayer coupling is favorable for all spins. The wave vector will be the largest where the interlayer coupling is not present. Here, with  $J_{\rm z}/J_{\rm xy}=0$ the wave vector is the same as a single layer system, and goes to zero where the interlayer coupling is much larger than the DMI. 

\begin{figure}
\includegraphics[width=0.9\linewidth]{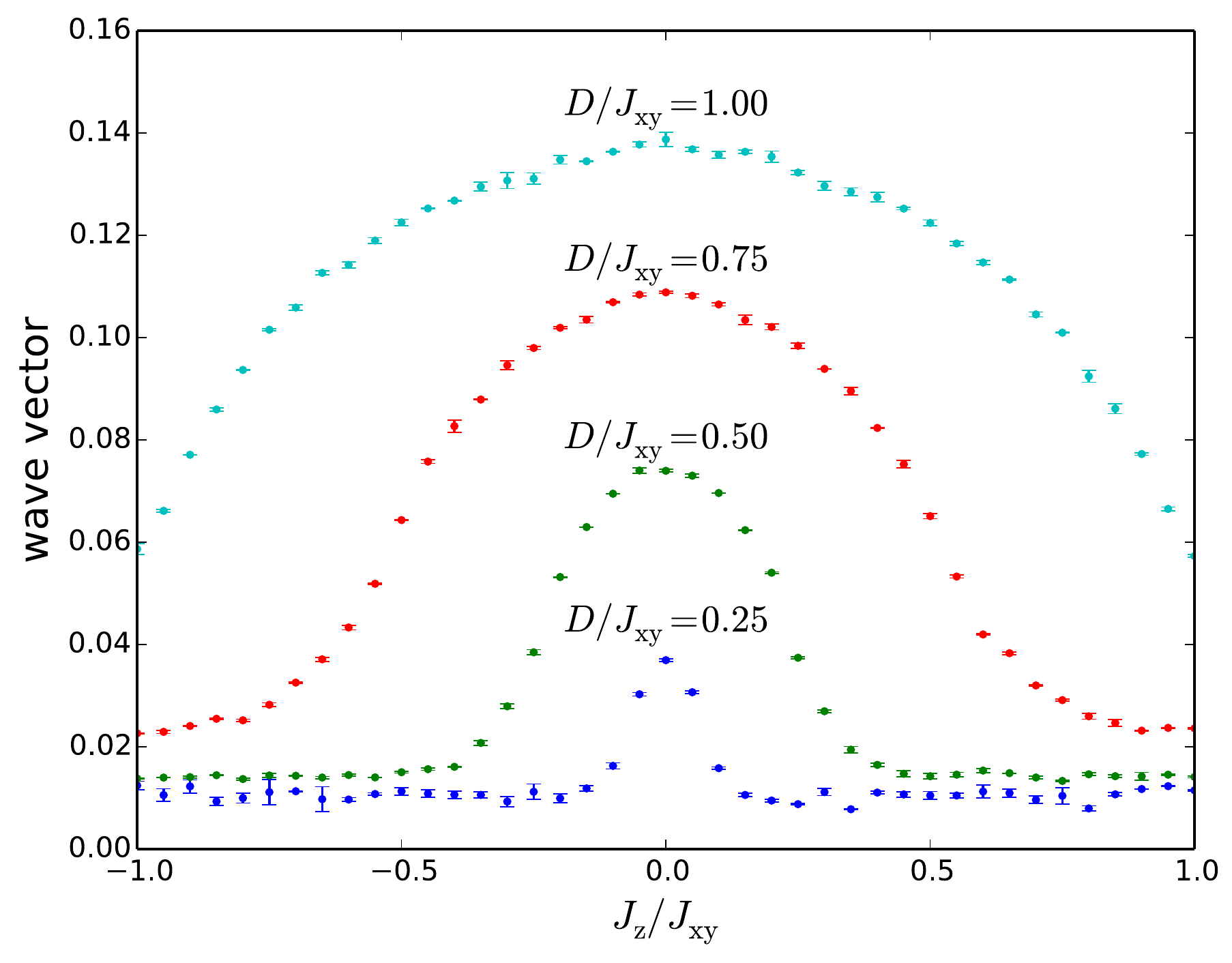}
\caption{Wave vector for various values of DMI plotted for a range of interlayer coupling for a 5 layer system. A small wave vector corresponds to a larger spiral wave length.}
\label{fig:wavevector}
\end{figure}

\begin{figure}
\includegraphics[width=0.9\linewidth]{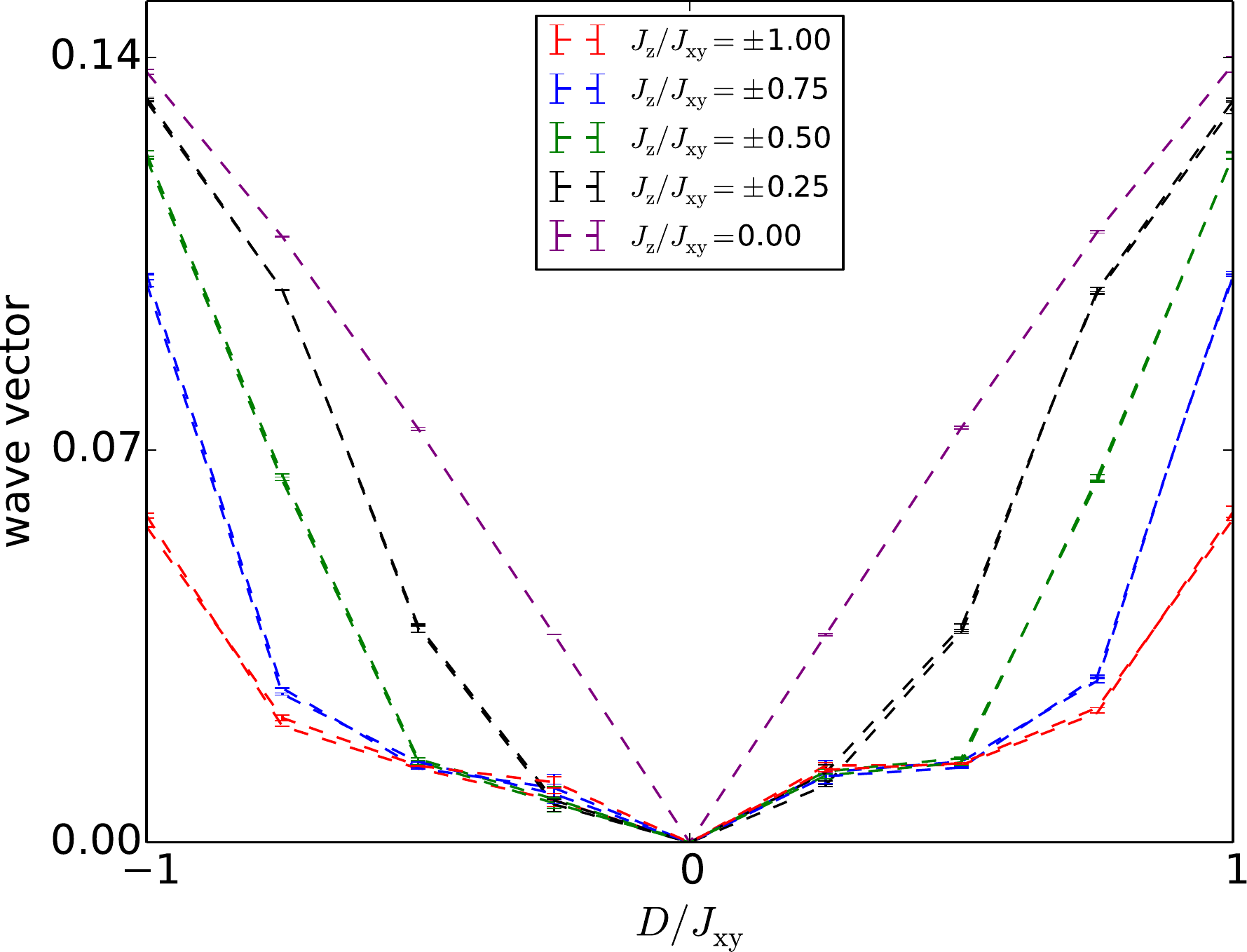}
\caption{Wave vector for various values of interlayer coupling plotted for a range of DMI for a 5 layer system corresponding to \cref{fig:wavevector}. It is visible that the linear behavior of the wave vector is affected by the interlayer coupling.}
\label{fig:wavevector_DMI}
\end{figure}

In \cref{fig:wavevector} we show the wave vector for constant DMI, $D/J_{\rm xy}$ ranging from 0 to 1, and varying interlayer exchange. A big variation in the wave vector is found between small and large interlayer exchange. The wave vector for large interlayer exchange is around one-fifth the size of the wave vector for small interlayer exchange. The drop off between these two cases corresponds to the phase transition in the phase diagram between phase II and III. Thus we see that phase II has a larger wave vector than phase III. Also here there is no clear distinction visible between the ferromagnetic and antiferromagnetic interlayer coupling. 
To examine the distance between the different cases of $D/J_{\rm xy}$ in \cref{fig:wavevector} we plot the wave vector dependence on the DMI for different values of interlayer coupling $J_{\rm z}/J_{\rm xy}$ in \cref{fig:wavevector_DMI}. Here we see that the relation between DMI and wave vector without any interlayer coupling is linear, but for increasing interlayer coupling this linearity is not found since the wave vector is dependent on the different phases of the system. 

Since the wave vector is influenced by finite size effects we used a finite size scaling to determine the true wave vector. For this, simulations were preformed for systems with system size 64x64, 128x128 and 256x256 and we extrapolated the wave vector linearly in $1/L$. 
An example of a plot with different system sizes and the resulting true wave vector is shown in \cref{fig:evenlayers}. In the inset we show one of the data fits we used to determine the wave vector. The wave vector is determined as the average distance to the origin for all pixels in a 2 dimensional Fourier transform that are more then 5 standard deviations above the mean of all pixels in a layer, divided by the system size.  

\subsection{Influence of the number of layers}\label{n_layers}
\begin{figure}
\includegraphics[width=0.9\linewidth]{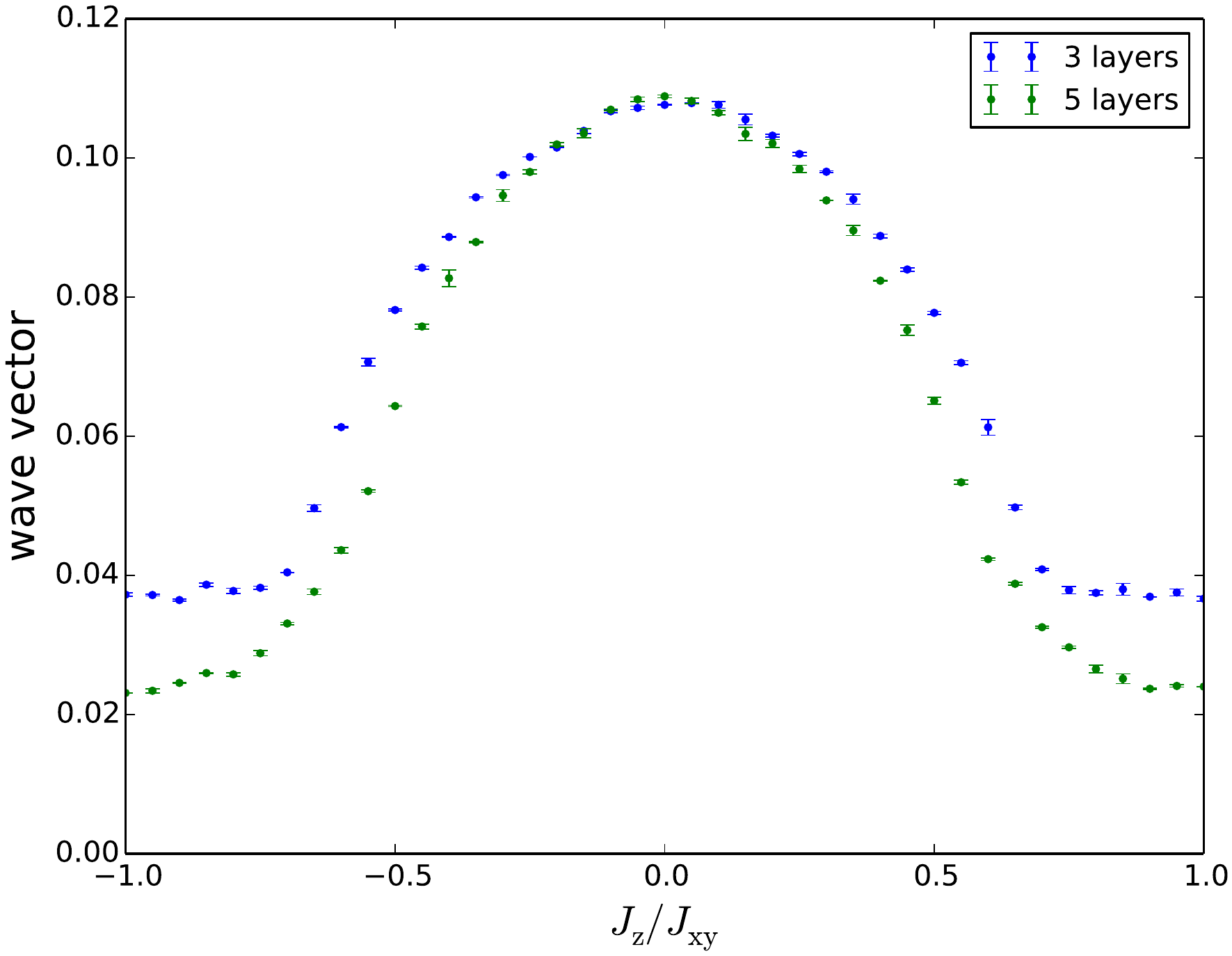}
\caption{Comparison of the wave vectors of a system with 3 layers and a system with 5 layers. Both layers are plotted for a DMI of $D/J_{\rm xy}=0.75$. A drop of between 2.6 and 3.2 is visible in the three layered system, and between 3.3 and 5.3 in the five layered system.}
\label{fig:difflayers}
\end{figure}

We investigated the wave vector for systems with different numbers of layers. See \cref{fig:difflayers} for a comparison between two systems with three and five layers. It is visible that the drop off is larger for a larger number of layers. Examining the ratio between the long wavelength in the DMI dominant phase and the shorter wavelength in the interlayer coupling dominant phase we find two effects. Firstly, this ratio is roughly 3 for a system with three layers and around 5 for a system with five layers. 
Our hypothesis is that the sum of the DMI divided by the number of layers gives a net DMI for strong interlayer coupling. 
At $J_{\rm z}/J_{\rm xy} = 0$ the layers are completely decoupled and behave as individual layers. Here the net DMI in each individual layer is exactly $D/J_{\rm xy}$. As we increase $J_{\rm z}$, the interlayer coupling leads to a `dilution' of the DMI and at very large interlayer coupling the system behaves as a system with a single DMI value, given by the sum of the DMI in each individual layer divided by the number of layers. As the number of layers increases, the DMI gets divided by an increasing number, Therefore leading to a decrease in the net total DMI and a reduction of the observed wave vector.
Because of the alternating DMI the total sum of the DMI is $D$ leading to a net DMI $D_{\rm net} = D/3$ for three layers and and $D_{\rm net} = D/5$ for five layers. This is only a rough approximation which works within 10\% of the ratio.
Secondly, the precise ratio of the layers themselves shows a pattern. There is a symmetry, the two outermost layers have the same ratio, while the inner layer has a different ratio. For example, in a 5 layered system the second and fourth layer have the same ratio and there is a symmetry around the middle layer. This hints to a more rich structure between the layers. We show the ratios in \cref{table}
, for which we used a system with $J_{\rm z}/J_{\rm xy}=\pm 1$ for the wave vector of a interlayer coupling dominating phase, and the values at $J_{\rm z}/J_{\rm xy}=0$ as the DMI dominating phase.

\begin{table}[]
\begin{tabular}{l|llllll}
\hline
$D/J_{\rm xy}$	& \textbf{-0.75} & \textbf{-0.50} & \textbf{-0.25} & \textbf{0.25} & \textbf{0.50} & \textbf{0.75} \\ \hline
Layer 1 		& 2.88           & 2.96           & 2.66           & 2.82          & 3.14          & 2.88          \\
Layer 2 		& 3.00           & 3.04           & 2.69           & 2.75          & 3.24          & 3.01          \\
Layer 3 		& 2.91           & 2.95           & 2.60           & 2.74          & 3.13          & 2.89          \\ \hline
\end{tabular}
\caption{Table of the ratio between the wave number in the DMI dominating phase III and phase II in a three layer system. Columns are for $D/J_{\rm xy}$, and rows for the different layers in a system. This is for a 3 layer system. The ratio is determined for $J_{\rm z}/J_{\rm xy}=\pm1$ and $J_{\rm z}/J_{\rm xy}=0$. The results for $D/J_{\rm xy}=\pm1$ are omitted in this table because for this strong the interlayer coupling must be stronger than the intralayer coupling to reach phase II. }
\label{table}
\end{table}

\begin{table}[]
\begin{tabular}{l|llllll}
\hline
$D/J_{\rm xy}$ 	& \textbf{-0.75} & \textbf{-0.50} 	& \textbf{-0.25} 	& \textbf{0.25} 	& \textbf{0.50} 	& \textbf{0.75} \\ \hline
Layer 1 		& 4.47           	& 4.65           	& 3.37           	& 3.25          	& 5.21          	& 4.57          \\
Layer 2 		& 4.65           	& 4.77           	& 3.38           	& 3.34          	& 5.31          	& 4.60          \\
Layer 3 		& 4.56           	& 4.68           	& 3.40           	& 3.29          	& 5.28          	& 4.59          \\
Layer 4 		& 4.66           	& 4.77           	& 3.37           	& 3.27          	& 5.34          	& 4.63          \\
Layer 5 		& 4.51           	& 4.64           	& 3.44           	& 3.24          	& 5.21          	& 4.51          \\ \hline
\end{tabular}
\caption{Table of the ratio between the wave number in the DMI dominating phase III and phase II in a five layer system. Columns are for $D/J_{\rm xy}$, and rows for the different layers in a system. This is for a 5 layer system. Our approximation of  $D_{\rm net}$ breaks down at $D/J_{\rm xy}=\pm0.25$ where the ratios are much smaller. The ratio is determined for $J_{\rm z}/J_{\rm xy}=\pm1$ and $J_{\rm z}/J_{\rm xy}=0$. The results for $D/J_{\rm xy}=\pm1$ are omitted in this table because for this strong the interlayer coupling must be stronger than the intralayer coupling to reach phase II. }
\label{table}
\end{table}


\subsection{Results for even number of layers} \label{secEven}
\begin{figure}
\includegraphics[width=0.9\linewidth]{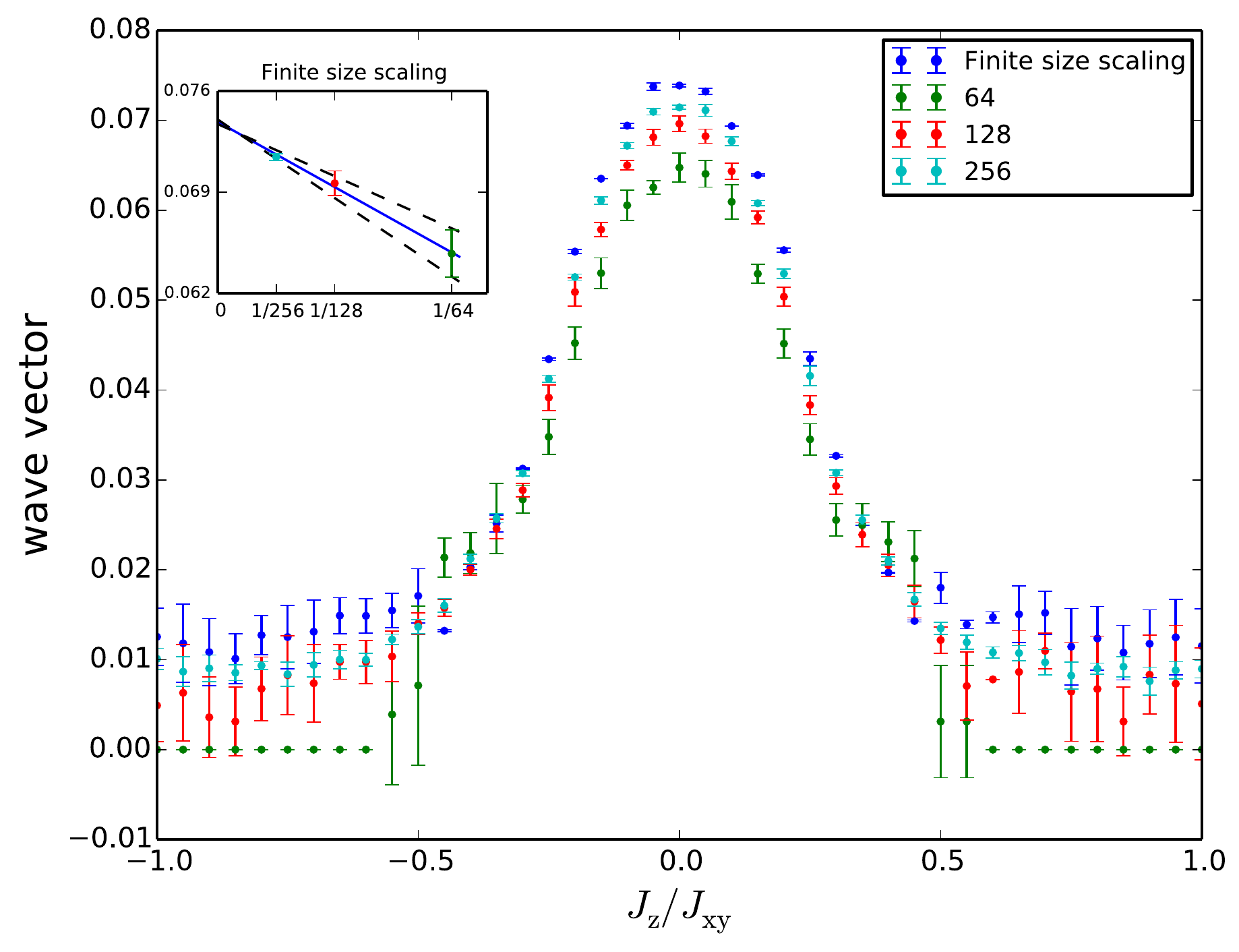}
\caption{Wave vector plotted against interlayer coupling $J_{\rm z}/J_{\rm xy}$ for a system with 4 layers and a DMI of $D/J_{\rm xy}=0.50$. Different system sizes 64x64, 128x128 and 256x256 are plotted, and a finite size scaling made with a linear fit in $1/L$ is added. The inset shows the finite size scaling for $J_{\rm z}/J_{\rm xy} = 0$. }
\label{fig:evenlayers}
\end{figure}

When a thin ferromagnet with local inversion asymmetry has an even number of layers the sum of the total DMI will be zero, i.e. $D_{\rm net}$ will be zero.
Thus in the ferromagnetic dominant phase we expect to observe no spin spirals. This is indeed the case. In \cref{fig:evenlayers} we plot the wave vector of a system with 4 layers. In the DMI dominant phase we still find spin spirals with a wave vector of $0.07$ inverse lattice spacings, which is comparable to the odd layered case. However, in the interlayer coupling dominate phase the wave vector drops to the predicted $0$. 
In \cref{fig:evenlayers} it is visible that this is indeed the case for a system size of 64x64 spins, but for larger system sizes the wave number does not drop to $0$ exactly. Further examining snapshots of the systems we see that in the interlayer dominated region no spin spirals are present, as expected. However, we observe localized spin textures, a region where the spins are oriented differently from the rest of the polarized system. These are formed by thermal fluctuations around the nucleation temperature and are stabilized by the DMI which is still present. 
By comparing these textures to uniform magnetized textures we found that these artifacts have a higher energy and are thus not representative for the ground state. The error bars for these structure also indicate the volatility of these textures and indicate that they are not the ground state. 

\section{Conclusion, discussion and outlook} \label{secConcl}
In conclusion, we presented results on the behavior of spin textures in ferromagnets with a local inversion asymmetry between the layers which leads to alternating DMI for consecutive layers. We found strong effects for the chiral spiral textures originating from the DMI. Furthermore, we were able to distinguish three different phases: the polarized phase, DMI-dominant phase and the interlayer coupling-dominant phase. The interlayer coupling-dominant phase has shown different behavior for the number of layers in a system by influencing the wavelength of the spirals, and also shown a strong difference between odd en even number of layers in a system. 

In this paper we used a minimal model. Future research can be focussed to include additional magnetic effects known to influence textures, such as anisotropy, dipole-dipole interactions and an external magnetic field. 
While the finite size scaling of our results gives a good indication of the expected wave number and a maximum system size of 256x256 spins is still efficient to simulate, more certainty in the wave number is possible with larger system sizes. 
More computations can also be used to test our hypothesis of the net DMI that occurs in coupled layers. A higher number of layers should continue the trend reported in this paper. 
Furthermore, the phase diagram in this paper gives a direction to where the different phases occur, a more detailed version can be achieved by simulations for more parameters. 

Our results shown here demonstrate that rich magnetic structures can be obtained through local DMI engineering.
First, as a useful method to measure the Dzyaloshinskii-Moriya interaction in \vdwfm\ or multilayer ferromagnets with a broken local inversion symmetry. When the number of layers and coupling between the layers is known, the DMI can be determined by looking at the spiral wave vector in the upper layer.
Second, possible applications that need the ability to tune spiral wave length or even completely turn off spin spirals. This can be achieved by changing the coupling between the layers e.g. through applying pressure a Van der Waals material \cite{Song2019,Li2019} or by changing the spacing layer in a RKKY coupled metallic stack. 
Third, our results serve as a prelude to investigating skyrmions in Van der Waals magnets. Moreover, an addition of an external magnetic field should lead to the formation of skyrmions. 

R.D. is member of the D-ITP consortium, a program of the Dutch Organization for Scientific Research (NWO) that is funded by the Dutch Ministry of Education, Culture and Science(OCW). This project has received funding from the European Research Council (ERC) under the European Union?s Horizon 2020 research and innovation programme (grant agreement No. 725509). This work is part of the research program Skyrmionics - towards skyrmions for nanoelectronics, which is financed by the Dutch Research Council (NWO). M.H.D.G. acknowledges the financial support of NWO (VENI 15093).

\bibliography{citaties}

\end{document}